

\input mn
\input epsf


\let\sec=\section
\let\ssec=\subsection


\def\bigstrut{\vrule width0pt height0.35truecm}
\font\japit = cmti10 at 11truept
\def\ss{\scriptscriptstyle\rm}
\def\ref{\parskip =0pt\par\noindent\hangindent\parindent
    \parskip =2ex plus .5ex minus .1ex}
\def\gs{\mathrel{\lower0.6ex\hbox{$\buildrel {\textstyle >}
 \over {\scriptstyle \sim}$}}}
\def\ls{\mathrel{\lower0.6ex\hbox{$\buildrel {\textstyle <}
 \over {\scriptstyle \sim}$}}}
\newcount\equationo
\equationo = 0

\newcount\fred
\fred=0

\def\outeqn#1{\the #1}

\def\leftdisplay#1$${\leftline{$\displaystyle{#1}$
  \global\advance\equationo by1\hfill (\the\equationo )}$$}
\everydisplay{\leftdisplay}

\def\eol{\hfill\break}

\def\mum{\mu{\rm m}}
\def\msunyr{\,h^{-2}{\rm M}_\odot\rm yr^{-1}}

\def\japfig#1#2#3#4{
\ifnum #2 = 0
\beginfigure{#1}
\epsfxsize=8.2cm
\noindent\centerline{\epsfbox[28 186 488 590]{starb_#3.eps}}
\fi
\ifnum #2 = 1 
\beginfigure{#1}
\epsfxsize=8.2cm
\noindent\centerline{\epsfbox[60 210 515 600]{starb_#3.eps}}
\fi
\ifnum #2 = 2 
\beginfigure{#1}
\epsfxsize=8.2cm
\noindent\centerline{\epsfbox[0 0 471 474]{starb_#3.eps}}
\fi
\ifnum #2 = 3 
\beginfigure{#1}
\epsfxsize=8.2cm
\noindent\centerline{\epsfbox[10 132 488 618]{starb_#3.eps}}
\fi
\ifnum #2 = 4 
\beginfigure*{#1}
\epsfxsize=11.5cm
\noindent\centerline{\epsfbox[0 0 379 578]{starb_#3.eps}}
\fi
\ifnum #2 = 5 
\beginfigure{#1}
\epsfxsize=8.2cm
\noindent\centerline{\epsfbox[38 124 522 612]{starb_#3.eps}}
\fi
\ifnum #2 = 6 
\beginfigure*{#1}
\epsfxsize=13.5cm
\noindent\centerline{\epsfbox[0 0 452 218]{starb_#3.eps}}
\fi
\caption{%
{\bf Figure #1.}
#4
}
\endfigure
}



\def\aj{AJ}

\def\apj{ApJ}

\def\mn{MNRAS}

%

\pageoffset{-0.8cm}{0.2cm}




\begintopmatter  

\vglue-2.2truecm
\centerline{\japit Accepted for publication in Monthly Notices of the R.A.S.}
\vglue 1.7truecm

\title{Starburst galaxies and structure in the submillimetre 
background towards the Hubble Deep Field}

\author{J.A. Peacock$^1$, 
M. Rowan-Robinson$^2$,
A.W. Blain$^{3,6}$,
J.S. Dunlop$^1$, \eol
A. Efstathiou$^2$,
D.H. Hughes$^{1,7}$, 
T. Jenness$^4$,
R.J. Ivison$^5$,
A. Lawrence$^1$, \eol
M.S. Longair$^3$,
R.G. Mann$^{2,1}$,
S.J. Oliver$^{2,8}$,
S. Serjeant$^2$
}

\affiliation{$^1$Institute for Astronomy, University 
of Edinburgh, Royal Observatory, Blackford Hill, Edinburgh EH9 3HJ, UK \bigstrut\eol
$^2$Astrophysics Group, Imperial College, Blackett Laboratory,
Prince Consort Road, London SW7 2BZ, UK \eol
$^3$Cavendish Astrophysics Group, Cavendish Laboratory, 
Madingley Road, Cambridge CB3 0HE, UK \eol
$^4$Joint Astronomy Centre, 660 N. A'ohoku Place, Hilo, Hawaii 96720, USA \eol
$^5$Department of Physics \& Astronomy, University College London, Gower Street, London WC1E 6BT, UK \eol
$^6$Institute of Astronomy, University of Cambridge, Madingley Road, Cambridge CB3 0HA, UK \eol
$^7$Instituto Nacional de Astrofisica, Optica y Electronica (INAOE), Apartado Postal 51 y 216, 
72000 Puebla, Pue., Mexico \eol
$^8$CPES, University of Sussex, Falmer, Brighton BN1 9QH, UK
}

\shortauthor{J.A. Peacock et al.}

\shorttitle{Structure in the submillimetre background towards the HDF}


\abstract{%
We use an 850-$\mum$ SCUBA map of the HDF to study the
dust properties of optically-selected starburst galaxies at high redshift.
The optical/IR data in the HDF allow a photometric redshift
to be estimated for each galaxy, together with an estimate
of the visible star-formation rate. The 850-$\mum$ flux density of
each source provides the complementary information: the
amount of hidden, dust-enshrouded 
star formation activity. Although the
850-$\mum$ map does not allow detection of the majority of individual
sources, we show that the galaxies with the highest
UV star-formation rates are detected statistically,
with a flux density of about $S_{850}=0.2$~mJy for 
an apparent UV star-formation
rate of $1\msunyr$.
This level of submillimetre output indicates that the
total star-forming activity is on average a factor of approximately 6 times
larger than the rate inferred from the UV output of these galaxies.
The general population of optical starbursts is then
predicted to contribute at least 25\% of the 850-$\mum$ background.
We carry out a power-spectrum analysis of the map, which yields
some evidence for angular clustering of the
background source population, but at a level lower than
that seen in Lyman-break galaxies.
Together with other lines of argument, particularly from the NICMOS HDF data,
this suggests that the 850-$\mum$ background originates over an extremely wide range
of redshifts -- perhaps $1 \ls z \ls 6$.
}


\maketitle  

\sec{INTRODUCTION}

The observational ingredients for an understanding of galaxy
formation are arguably now largely in place. In particular, the
Lyman-break technique has allowed the selection
of galaxies over the redshift range $2\ls z \ls 5$
(Steidel et al. 1996). The combination of ground-based and HST
data has allowed such samples to probe a wide range of
optical/UV luminosities, so that the statistical properties
of the high-redshift galaxy population are increasingly
well-known. This knowledge is conveniently summarized in the
form of the global star-formation rate as a function of redshift.
This function rises steeply between $z=0$ and $z=1$
(Lilly et al. 1996); initial indications from the HDF were that
there was then a maximum, followed by a decline at $z\gs 2$
(Madau et al. 1996).
More recent Lyman-break data, however, have cast doubt on this
claim: it now seems more likely that the UV luminosity density
remains approximately constant for $2\ls z \ls 5$ (Steidel et al. 1999).
Nevertheless, since there is little cosmological time at
$z > 5$, these results are often taken to suggest that we have seen almost all
the star-formation that ever occurred. Provided we can understand
the masses of the galaxies that are involved (as is starting
to be possible though Balmer-line spectroscopy; Pettini et al. 1998),
the entire history of the galaxy-formation process is potentially
open to view.

The main objection to this optimistic picture is well known:
star formation in the local universe is often accompanied by 
dust. There is thus the possibility that a large, or even dominant,
component of the star-forming history is not revealed by
optical/UV data.
It has long been clear that this question would only be settled
by studies in the far-IR or sub-mm bands, and great progress is
now being made in this area. The total amount of energy released
from young stars and reprocessed into emission from cool dust
is constrained by the detection of the background
radiation at $\lambda\gs 150 \,\mum$ (Puget et al. 1996; 
Schlegel, Finkbeiner \& Davis 1998;
Fixsen et al. 1998; Hauser et al. 1998). The resolution of this
background into point sources has been made possible with the
SCUBA imager on the James Clerk Maxwell Telescope
(Holland et al. 1999). A number of groups have used this facility
in order to map the sub-mm sky down to an rms sensitivity
below 1~mJy at $850\,\mum$
(Smail et al. 1997; Hughes et al. 1998; Blain et al. 1999;
Barger, Cowie \& Sanders 1999; Eales et al. 1999).

The observations of Hughes et al. (hereafter H98) are the deepest of existing
datasets, and also have the advantage that they cover the area
of the northern Hubble Deep Field (Williams et al. 1996). 
The galaxy population in this
region of the sky has been studied in great detail, and so
there is the potential to extract considerable information
from the sub-mm data. The initial study by H98 considered
only possible optical counterparts for the brightest 5 850-$\mum$
sources. Here, the intention is to dig much more deeply into
the background population. The optical HDF data tell us
the location of a large number of starburst galaxies, and it
should be possible to measure in a statistical way their
sub-mm emission. In this way, we will learn about the
fraction of the star-forming activity in these galaxies
that is hidden from view. The existence of photometric
redshifts means that we can do this as a function of redshift;
this measurement is critical, because it will tell us
whether or not to believe the indications from the optical/UV data
that most of the star-formation in the universe has been seen.

In fact, there are already indications that a
large fraction of the star-forming activity in the universe
is not visible in the UV. 
From the first general detection of luminous infrared galaxies
(e.g. Soifer et al. 1987), it has increasingly become clear that 
the majority of the star-forming activity at low redshift is
extinguished by dust (e.g. Meurer et al. 1997).
The same appears to be true at high redshift: for example,
the brightest few sources in the
HDF 850-$\mum$ map of H98 have very weak or no optical
counterparts. If their sub-mm emission arises from
star formation, all but of order 1\% of the activity is hidden.
This is direct evidence that the 850-$\mum$ background contains
a substantial contribution from ultraluminous infrared galaxies (ULIRGs),
as has been found by a number of other authors
(e.g. Barger et al. 1999; Lilly et al. 1999).
Even where starburst galaxies are visible in the UV,
measures of the star-formation rate (SFR) based on Balmer
emission lines often gives rates up to 10 times larger
than would be inferred from the UV continuum alone
(Pettini et al. 1998). These latter corrections are often
used to correct the observed UV luminosity density and thus
estimate the total density of star-formation in the universe.
The results of H98 show that this is dangerous, since there
can also be an additional component of star-formation that is
so heavily embedded that it fails to show up either
in the UV or in the Balmer lines. The only way to
detect such a component is through the sub-mm emission,
which is the approach of this paper. The structure of the
paper is as follows: section 2 summarizes the optical HDF
data and shows how to predict sub-mm emission from the HDF
galaxies as a function of redshift, to within an unknown
proportionality factor; section 3 compares these predictions
to the observed 850-$\mum$ data and shows that there is
evidence that the 850-$\mum$ emission originates over
the broad redshift range $1 \ls z \ls 6$;
section 4 attempts to test this picture by measuring
the clustering of the sub-mm emission in the HDF map.

\japfig{1}{0}{photz}
{Comparison of photometric redshifts between the
Imperial College group (Rowan-Robinson 2000) 
and Fernandez-Soto, Lanzetta \& Yahil (1999)
Solid points show spectroscopic redshifts (adopted
for the $x$ axis where known).}

\sec{STARBURSTS IN THE HDF}

\ssec{Photometric redshifts}

One great virtue of the HDF data is their wide wavelength
coverage. This has allowed several groups to 
estimate redshifts and spectral types via a
variety of template-fitting strategies.
Generally, these work impressively well
(see e.g. Hogg et al. 1998 and refs therein).
The catalogues
of Fernandez-Soto, Lanzetta \& Yahil (1999; FLY)
have been particularly successful, as they incorporate
$J$$H$$K$ data.
For the present work, we use a set of
photometric redshifts constructed according to a similar philosophy
at Imperial College (Rowan-Robinson 2000).
A comparison of their results is shown in Fig. 1;
generally, there is excellent agreement. The IC
database also provides redshift estimates for fainter galaxies,
beyond the reach of near-IR data. Inevitably, these estimates
will be less accurate, but are still worth including.

\ssec{UV star-formation rates}

Following Madau et al. (1996),
we estimate the star-formation rate from the luminosity
at a rest-frame wavelength of 1500\AA:
$$
{{\rm SFR} \over {\rm M}_\odot \rm yr^{-1}} =
{L_\nu ({\rm 1500\AA}) \over 10^{21.0}\, \rm W\, Hz^{-1}}.
$$
A number of slightly different figures for this conversion
can be found in the literature (e.g. Pettini et al. 1998;
Steidel et al. 1999); the uncertainty in the appropriate
conversion factor is however small compared to the
uncertain degree of extinction.
For redshifts $1 \ls z \ls 4.4$, this wavelength lies within
the range of the WFPC2 data; for other redshifts, 
a small degree of extrapolation is required.

The spectral index can also be evaluated (using the two
bands closest to 1500\AA): $L_\nu \propto \nu^{-\alpha}$.
Both the SFR and $\alpha$ require the data to be corrected for the
effects of intergalactic absorption. We use the corrections
given in Fig. 1 of Madau et al. (1996), which may conveniently
be approximated by
$$
{f_{\rm obs} \over f_{\rm emit}} =
\exp \left[ -(z/z_c)^{7.0} \right],
$$
with $z_c=(2.50,3.75,4.65)$ for the $(U_{300},B_{450},V_{606})$
wavebands.
For very high redshifts, these corrections become sufficiently large that they cannot
be made reliably (given the uncertainty in redshift). Therefore, for
$z>4.5$, we assume a spectral index of zero and work from the
observed magnitude at $I_{814}$.

Dust-free starbursts would normally be expected to have 
$\alpha\simeq 0$, whereas many of the HDF galaxies have 
redder spectra.
Pettini et al. (1998) argue that this reddening is due to
dust. They show that the star-formation rates for 
$z\simeq 3$ starbursts as inferred from Balmer lines are
often much higher than those inferred from the UV flux,
by up to a factor 10. Moreover, they claim that the reddest
galaxies are those for which the correction factor is largest.
The spectral indices could be used to 
produce corrected star-formation rates, but we shall
not do this; the aim of this paper is to use the
sub-mm emission to infer the total amount of hidden
star formation, which can then be compared with the activity that is
directly visible in the UV.

\japfig{2}{1}{sfrz}
{UV SFR against redshift. The raw SFR values are low
in all cases, with only a handful of objects exceeding
$5\msunyr$.}

Fig. 2  shows the inferred star-formation rate 
for all HDF galaxies with photometric redshifts (assuming
$\Omega=1$ unless otherwise stated), plotted
against redshift.
The uncorrected rates are low, with only 5 galaxies
exceeding $5\msunyr$. 
We see that the most active galaxies
are found rather uniformly over the
redshift range $1.5 \ls z \ls 5$.
To the extent that sub-mm flux density is a weak function
of redshift at high $z$, this suggests that we would expect
a similar redshift distribution for 850-$\mum$ sources.

\ssec{Dust models and 850-micron emission}

In order to make predictions of the 850-$\mum$ flux density,
we use a simple single-temperature grey-body dust model, with
$T=50$~K and an emissivity proportional to $\nu^\beta$ with
$\beta=1.5$. The luminosity at rest wavelength $\lambda_0$
is then
$$
{L_\nu \over \rm W\, Hz^{-1}} = 10^{23.0}h^{-2}\,
\left({{\rm SFR}\over \msunyr}\right)\,
{y^{-(3+\beta)}\over \exp(1/y)-1},
$$
where $y=\lambda_0/288\,\mum$.
The coefficient of proportionality was set by comparing with
models for Arp220 (Hughes \& Dunlop 1999); for a star-formation
rate of $75 \msunyr$, we normalize to
an 850-$\mum$ flux density of 2.5~mJy at $z=3$.
As is well known, such models predict a sub-mm
flux density that is very nearly independent of
redshift. For $z\gs 1.5$, our assumptions give
an SFR of approximately $30 \msunyr$ for a 1~mJy source.

Of course, the sub-mm luminosity is a not a measure of the
total star-formation rate; rather, it measures the
fraction that is heavily embedded. One way to deal
with this would be to follow Pettini et al. (1998)
and scale the UV SFR
by a factor that depends on spectral index:
${\rm SFR} \rightarrow f(\alpha) {\rm SFR}$; the
sub-mm emission would then be proportional to the
corrected star-formation rate, times $[1-1/f(\alpha)]$.
However, such a procedure implicitly assumes a
simple geometry for the dust, and ignores
the possibility of an extra contribution from
heavily embedded star-formation regions. We shall
therefore prefer to derive the relation between the
visible and hidden SFRs empirically.

It is clear in advance that very substantial
extinctions will be required if the HDF starburst
galaxies are to be detectable in the sub-mm.
The rms noise level of the 850-$\mum$
map is 0.45~mJy, so we require a flux density of 1~mJy
for an individual detection, or several times 0.1~mJy
in order to be able to achieve a statistical detection.
The former figure corresponds to an SFR of about
$30 \msunyr$, whereas we have seen that the uncorrected
UV SFR figures generally lie more than a factor of 10
lower than this.
In order to measure the hidden component, we shall use the term
`predicted' 850-$\mum$ flux density to denote the
value that would be expected if the hidden SFR
equalled the rate deduced from the UV continuum:
$$
{\rm hidden\ SFR \over \rm visible\ SFR} =
{{\rm observed\ } S_{850} \over {\rm predicted\ } S_{850} }.
$$
Of course, this makes the assumption that the sub-mm
emission originates in starlight re-radiated by dust.
All that is directly measured is the total amount
of energy that is being re-radiated,
independent of its origin.

\japfig{3}{6}{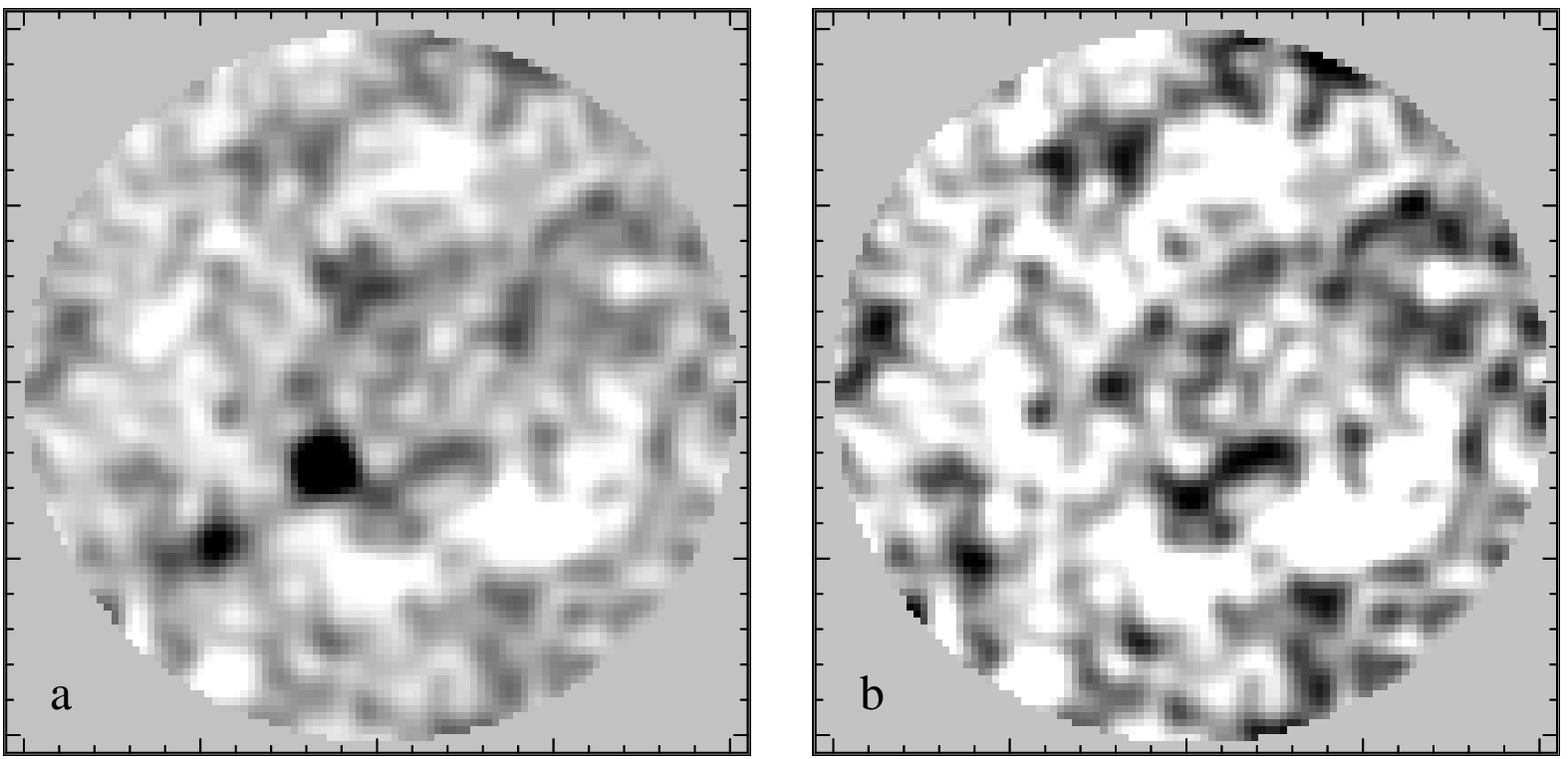}
{The 850-$\mum$ map of the HDF, with North to the
top and East to the left. Panel (a)
shows the raw map (in which 5 sources brighter than
2~mJy have been subtracted and restored with single Gaussian
beams, in order to reduce the effects of sidelobes); 
panel (b) shows the residual map with
these 5 sources removed. 
Both maps are 210$''$ square; 
the frame has small tick marks at 10$''$ intervals.
Black denotes high flux; the map is set
to zero beyond 100$''$ radius. 
Both maps have been scaled to remove the trend for
instrumental noise to increase with radius.}

\sec{STATISTICAL DETECTION OF OPTICAL STARBURSTS}

\ssec{Predicted SCUBA maps}

One direct way in which we can try to see if the general
starburst population contributes to the SCUBA data is
to make synthetic maps. 
Rather than compare directly to the
observed 850-$\mum$ map, we first construct a residual
map, subtracting the 5 brightest sources (to a limit of 2~mJy) studied
in H98. The rationale for this is that it is impossible
to predict accurately the brightest few sources, which
dominate the visual appearance of the map. On the other hand,
the general background emission is affected by many
objects, and we stand a better chance of seeing a
correlation between predictions and observations.
Fig. 3 shows the original and the residual map.

\japfig{4}{4}{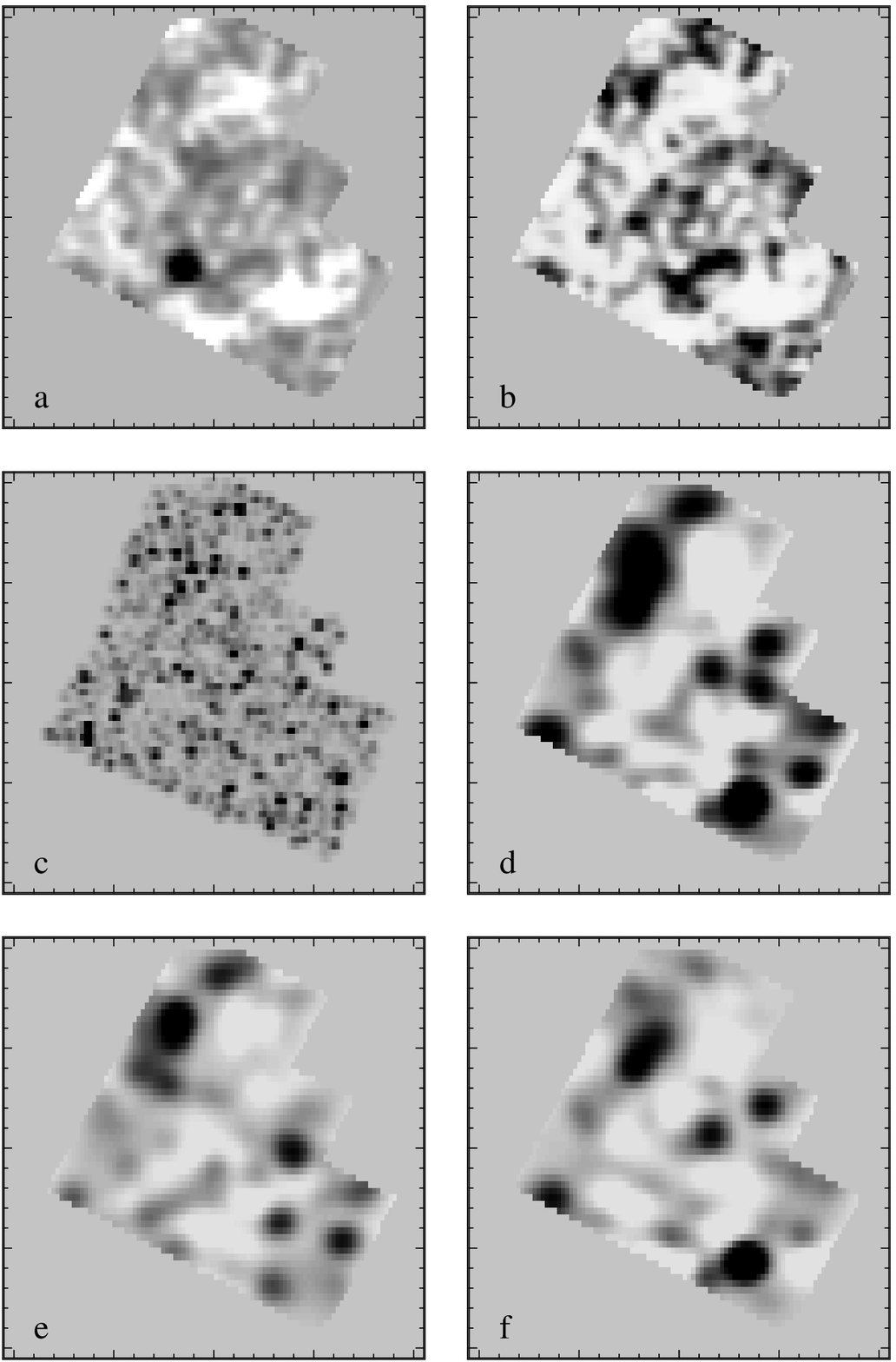}
{(a) The observed SCUBA 850-$\mum$ map, with North to the
top and East to the left, windowed over the
area of the HDF.
The frame plotted is $200''$ square; (b) the SCUBA residual map, with the 5
brightest sources ($>2$~mJy) subtracted; (c) the predicted 850-$\mum$ map
at high resolution; (d) the predicted 850-$\mum$ map
with the SCUBA beam; (e) and (f) show the same as
(d) for respectively $z<2.5$ and $z>2.5$.
Different greyscales are used in each case, to bring
out the detail of interest in a given map.
The frames show small tick marks at 10$''$ intervals.
}

The residual map (Fig. 3b) shows considerable structure.
There are a number of isolated regions of peak emission
around 2~mJy, which are further candidates for point
sources. Some of these clearly blend, such as the horizontal `worm'
structure just South of the map centre. This is an indication
that the image is starting to approach the confusion limit.
There is in addition a hint of a striped pattern, with
large-scale areas of low and high emission. This pattern
could be taken as a flat-fielding artefact, and a number
of reduction strategies were tried, with the conclusion that
it seemed to be a robust feature. 

We now ask for the predicted appearance of the 850-$\mum$
sky, based on the UV starbursts.
This is only possible over the sub-area of the 850-$\mum$ map that
overlaps the optical HDF, as shown in Figs 4a and 4b.
The predicted 850-$\mum$ map is shown in Fig 4c;
this represents a possible future view of the 
sub-mm sky, given an instrument with an effective
aperture a few times larger than the JCMT.
When convolved with the JCMT beam (including the effects of
chopping and nodding), the simulated (noise-free) map of 
Fig. 4d is produced.
As a match to the true sky, Fig. 4d has some successes
and some failures. It does reproduce the character of the large-scale
variations in emission. This encourages us to believe that
features such as the prominent `hole' to the bottom right
(WF chip 4) do indeed result from a general lack of sub-mm sources
in that region of the sky. 
Other features match less well, however: the predicted
top central bright feature is absent in reality;
there is no optical counterpart to the prominent horizontal
blend just South of the map centre.
This lack of a detailed match is entirely to be expected,
given that there is a large contribution to the map from 
ULIRGs, which cannot be predicted from the HDF UV data.

\ssec{Emission in redshift shells}

It is interesting to dissect the predicted sub-mm emission
as a function of redshift. It is entirely possible that
the relative amounts of hidden and visible star formation could
evolve as a function of 
redshift, causing one redshift band to dominate
more than predicted.
We therefore divided the map of Fig. 4d into the contributions from different
redshift slices, and looked to see how well each correlated with the true map;
the result is shown in Fig. 5.
Although the correlation signals from different
redshift shells are relatively noisy,
there is a quite impressive trend with redshift, with 
little mean correlation at
$z<1$, but a positive signal broadly distributed over the
higher-redshift bins -- especially
at $z>2.5$. Figs 4e \& 4f show the contributions to the
predicted sub-mm map, divided at this redshift.
This is direct evidence that, as expected from the SFR data
in Fig. 2, the HDF map receives contributions from a very wide range
of redshifts.
The UV starburst galaxies thus contribute to the 850-$\mum$
background in a similar
fashion to the ULIRG population, which has
a median redshift of between 2.5 and 3 (H98; Smail et al. 2000).

\japfig{5}{1}{xcorr}
{The cross-correlation coefficient between the 850-$\mum$ residual map and
the projected UV emission from different redshift shells.
The error bars were estimated by rotating the SCUBA
map relative to the optical HDF data.
Although no single redshift shell shows a strongly
significant cross-correlation signal, there is a significant
general level of correlation between the SCUBA data and the high-redshift
galaxy distribution, broadly distributed over the redshift
range $z>1$.}

\japfig{6}{2}{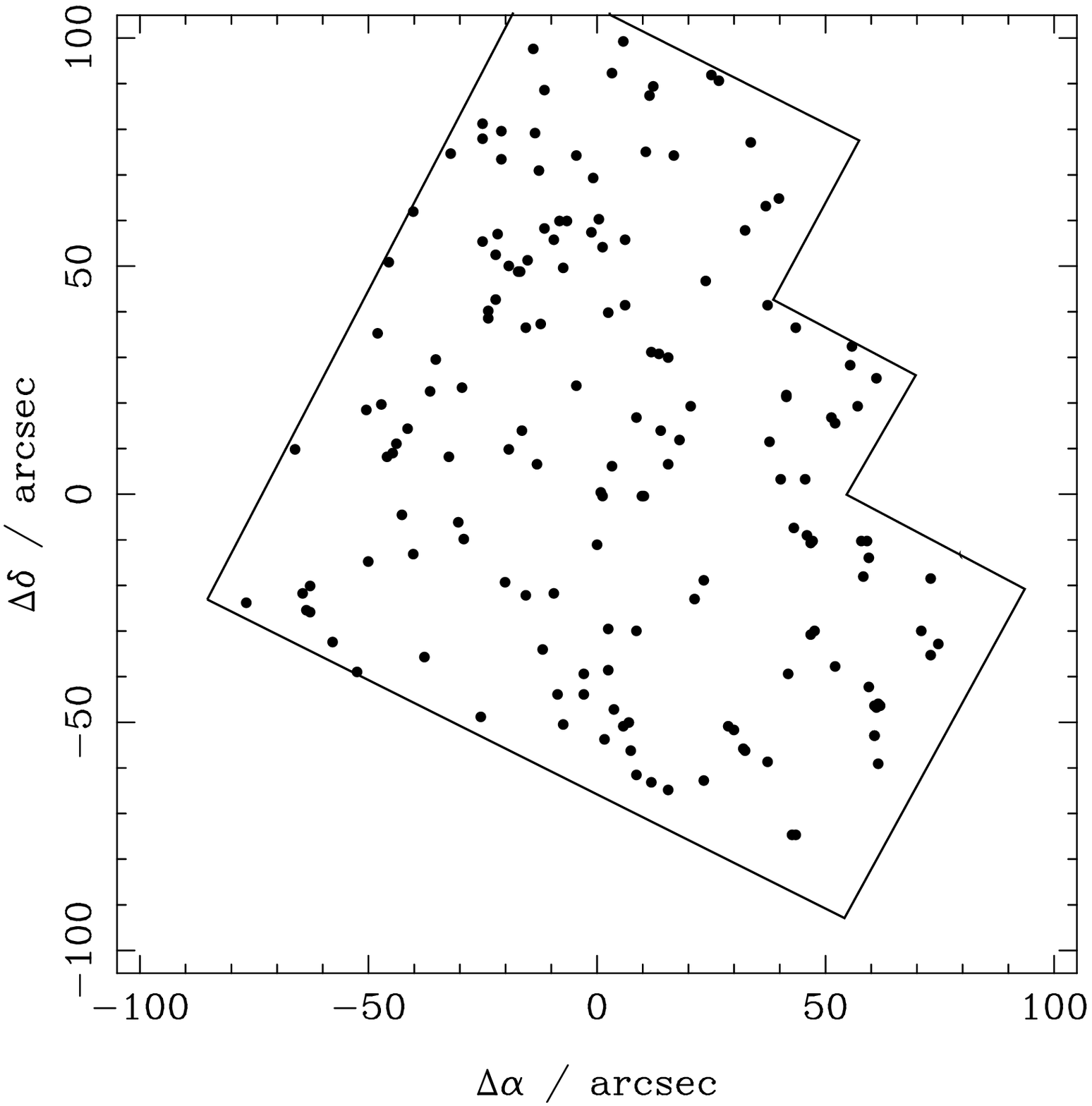}
{The brighter ($I_{814}<27$) starburst galaxies at $z>2.5$. There is
a striking `hole' in the distribution, centred on
$(x,y)=(40,-40)$, corresponding to the observed lack of
emission in the SCUBA map.}

Fig. 6 shows in detail the distribution of $z>2.5$ galaxies
(confined to the brighter objects, where the photometric
redshifts should be more robust). This is clearly
a very strongly clustered distribution, as found
by Steidel et al. (1998). Perhaps the most striking feature
is the lack of galaxies in the region where the
SCUBA map shows its deepest `hole'. 
Although we do not necessarily expect a very precise match between the observed
and the predicted $S_{850}$ for a given object, the existence of large-scale
features in a clustered galaxy distribution will induce
structure in the 850-$\mum$ map. We will return to this
point in section 4 below, where we quantify the degree
of clustering in the sub-mm background.

\ssec{The amount of hidden star formation}

Perhaps the most direct way of seeing if we are detecting the
effect of the UV starbursts on the 850-$\mum$ map is to
estimate a flux density for each galaxy, and look for
a correlation between predicted and observed flux densities.
We therefore take the value of the residual map at the position of
each UV source as the best estimate of its true 850-$\mum$
flux density.

\japfig{7}{5}{sz}
{Observed 850-$\mum$ flux density against redshift.
The bottom panel shows the raw values; the top
shows mean values with standard errors.}

\japfig{8}{5}{ssfr}
{Observed 850-$\mum$ flux density against UV SFR.
The bottom panel shows the raw values; the top
shows mean values with standard errors.}

These observed flux densities are plotted against redshift in Fig. 7.
This is a scatter diagram, with no correlation.
However, a very different picture emerges when we plot 850-$\mum$ flux density against UV SFR,
as in Fig. 8. Much of this plot is also a scatter diagram,
but closer inspection reveals a clear tendency for the
objects of highest star-formation rate to have positive
fluxes. There is a group of sources with
$S_{850}\gs 1$~mJy and ${\rm SFR} \gs 2\msunyr$ that stand
out as plausible real detections.
The parameters for these sources are listed in Table 1.
Averaging the data in bins yields mean values for the
flux density at a given SFR, and these results are also shown in
Fig. 8. This demonstrates that we are able to achieve a clear
statistical detection of 850-$\mum$ emission from optically
selected galaxies even at a level down to 0.1~mJy. 

It is interesting to contrast this result with the direct approach of pointed
observations of individual Lyman-break galaxies, which generally
yield upper limits of a few mJy (e.g. Chapman et al. 1999).
The statistical approach clearly needs a little thought, since
we cannot be certain of the detection of any individual
source. For example, many of the features on the HDF map
are probably blends of a number of sources, so is there not a
danger that we might be overestimating the mean flux from
any individual source by also counting its neighbours? 
In fact, there is no such bias, at least for a Poisson-distributed
field of sources (we consider the effect of clustering in Section 4
below). For a Poisson distribution, the expectation value
of the flux contributed by all neighbours is just the mean
flux on the map, which is zero owing to the chop-nod observing
mode. Clearly, the effect of blends contributes
a confusion-noise term to the general rms on the map, and
this makes it harder to detect any given source.
However, the instrumental noise dominates the rms in
practice, so this is a minor effect. Even though there
are a finite number of samples of noise on the map,
each randomly-placed source gives a statistically independent
sampling of this noise distribution, so the normal
statistics of mean and standard error can be applied.

The mean and standard error 
of the sub-mm flux density to SFR ratio inferred from the points 
in Fig. 8 is
$$
S_{850} / {\rm mJy} = 0.20\pm 0.04 \, {{\rm SFR}\over \msunyr}.
$$
This implies that the hidden star formation significantly exceeds
that visible in the UV.
A direct fit to the ratios of observed and predicted
flux densities gives
$$
{\rm hidden\ SFR \over \rm visible\ SFR} = 4.25 \pm 1.2.
$$

It is possible that this figure may be an underestimate.
Examining the entries in Table 1 in more detail, it is
apparent that there are a number of objects with
estimated $z>5$ for which (a) the IC and FLY estimates
disagree; (b) the observed flux densities
scatter around  zero. 
The estimated redshifts can tend to be unreliable at
the extremes, owing to detection in only a few bands,
so it is possible that the sample is corrupted by objects
whose apparent high SFR is due to an erroneously large
estimated $z$. If the above exercise is repeated, excluding
objects with $z_{\rm est}>5$, the figures increase by about $1\sigma$:
$$
S_{850} / {\rm mJy} = 0.25\pm 0.03 \, {{\rm SFR}\over \msunyr};
$$
$$
{\rm hidden\ SFR \over \rm visible\ SFR} = 6.85 \pm 1.53.
$$
An alternative approach is to repeat the entire exercise using only
the FLY redshift estimates; again the effect is to raise the
level of the statistically detected $S_{850}$ by very nearly $1\sigma$.

It is worth asking if these results are consistent with the
findings of Chapman et al. (1999), who carried out SCUBA
photometry of eight Lyman-break galaxies, obtaining only
one clear detection. The data in their Table 1 gives the
ratio between 850-$\mum$ flux density and SFR, which has
a typical rms error of 0.2 for an individual object.
If fact, the scatter of their values is about twice
this figure, apparently indicating a real scatter in properties.
It is therefore appropriate to evaluate an unweighted
mean and standard error, which gives
$$
S_{850} / {\rm mJy} = 0.13\pm 0.14 \, {{\rm SFR}\over \msunyr};
$$
this is a lower figure than our detection, but is
statistically perfectly consistent with it.

Taking the various estimates discussed above, we adopt
a best guess for the hidden-to-total ratio of
$$
{\rm hidden\ SFR \over \rm visible\ SFR} = 5 \pm 1.5.
$$
This is unfortunately a rather model-dependent
number, since it is sensitive to the spectral
shape assumed in the sub-mm: the predicted 850-$\mum$
flux density scales roughly as $T^{-3.5}$ for 
a fixed bolometric luminosity, so moderate changes to the
temperature can significantly alter the SFR
inferred from a given 850-$\mum$ measurement.
Nevertheless, this correction factor for the total SFR 
is similar to the mean correction inferred
by Pettini et al. (1998), and other pieces of evidence are consistent
with their results. Where reliable spectral indices can
be deduced for the sources in Table 1, they are large and
positive, ranging between 0.3 and 2.0. This is in agreement
with the claim by Pettini et al. that a red UV continuum
correlated with a high mean level of extinction.
Similar arguments were also advanced by Meurer, Heckman \& Calzetti (1999);
they used the observed UV spectra of HDF galaxies to predict
submillimetre fluxes, and their predicted results are in
broad agreement with our data.

However, one should not assume a physical picture in which a
foreground dust screen extinguishes a dust-free starburst.
In a sense, all star formation is dust-enshrouded in that
stars are born in dusty molecular clouds. Jimenez et al. (2000)
argue that it takes about 15~Myr for young massive stars
to burn away this dust and become visible. For
continuous star formation, the hidden-to-visible SFR
ratio is then really just a ratio of the time spent
enshrouded, to the remaining lifetime of the stars
once they are visible. Jimenez et al. (2000) argue that
this ratio should be about 6:1 for a Salpeter IMF, and
this is not inconsistent with our results.
It is only possible to exceed this ratio if all star-forming
regions are seen at an early phase, corresponding to a single
dominant starburst; in this case, the optical counterpart
will take the form of an ERO (e.g. Smail et al. 1999), or
will be completely undetectable.

\ssec{Contribution of Lyman-break galaxies to the background}

If the true volume density of star formation is about 6 times
what is visible is the UV, we need to be sure that this
is consistent with the counts and background at 850~$\mum$.
The background is related to the proper volume emissivity, $j_\nu$, as
a function of redshift by
$$
I_\nu(\nu_0) = {1\over 4\pi}\; {c\over H_0} \, \int j_\nu[(1+z)\nu_0]\, {dz\over (1+z)^{11/2}}
$$
(for $\Omega=1$; see e.g. chapter 3 of Peacock 1999).

Given the comoving redshift-dependent SFR density, it is easy to use this
expression to evaluate the background. We use the UV density 
estimated by Steidel et al. (1999), with an assumed
hidden-to-visible ratio of 5.
Integrating to $z=5$ gives a predicted background of
$$
\nu I^{850}_\nu ({\rm Lyman-break}) = 5.5 \times 10^{-10} \, {\rm W m^{-2}sr^{-1}},
$$
which is approximately identical to the total background 
($5\pm 2\times 10^{-10}\, {\rm W m^{-2}sr^{-1}}$) estimated by
Fixsen et al. (1998). 
This figure could be increased still further, by a factor
of about 1.5, if we had adopted the corrections
suggested by Steidel et al.
to allow integration over the whole UV luminosity function.
Alternatively, the estimate might be too high by a similar factor,
because the SFR density at $z\ls 2$ is deduced from luminosities
at 2800\AA, and these give systematically larger SFR figures for
the more luminous starbursts in this paper. Both these corrections
are within the statistical uncertainty on the SFR density.

Normal star-forming
galaxies thus clearly make a very significant contribution to the submillimetre background,
perhaps the dominant one.
At the bright flux densities investigated by H98,
all sources had hidden-to-visible SFR ratios that exceeded 60,
so ULIRGs dominate the background in this flux regime, 
which contributes roughly 30\% of the Fixsen et al. background.
Active galaxies may contribute up to a further 10\% (Almaini, Lawrence \& Boyle 1999;
Gunn \& Shanks 1999).
Taking the formal uncertainties on the background and on the
hidden-to-total ratio, the 95\% lower confidence limit on the
contribution of UV starburst galaxies to the background is 48\%,
which is just consistent with the other contributions. This
figure could be lowered in one of two ways: (i) 
if the true background were higher than estimated by Fixsen et al. (1998);
(ii) if this analysis has overestimated the 850-$\mum$ flux density
that corresponds to a given UV SFR. As discussed in section 4, the
only way in which the second effect might occur is as a result of clustering
in the background population. In this case, the effective sub-mm emission
might be overestimated by up to a factor 2. Even in this extreme case,
however, the UV starbursts must generate a minimum of 25\% of the
850-$\mum$ background. We note that this empirical conclusion is in
agreement with detailed attempts to fit the faint optical galaxy
counts with models that include a dust correction to the SED,
thus predicting also the contribution of normal galaxies to the
submillimetre background (Busswell \& Shanks 2000).

\def\min{\llap{$-$}}
\begintable*{1}
\caption{{\bf Table 1} The submillimetre properties of HDF
sources with the largest star-formation rates ($>2 \msunyr$).
True photometric redshifts are denoted by an asterisk.}

\centerline{
\vbox{
\offinterlineskip
\tabskip 0.45cm
\halign{\phantom{\llap{$A^\beta_\beta$}}
\hfil# & \hfil#\hfil & \hfil#\hfil & \hfil#\hfil & \hfil#\hfil & \hfil#\hfil & \hfil#\hfil & \hfil#\hfil & \hfil# & \hfil#\hfil & \hfil#\hfil \cr
Name & $I_{814}$ & $V_{606}$ & $B_{450}$ & $U_{300}$ & $\alpha$ & $z_{\ss IC}$ & $z_{\ss FLY}$ & SFR &
 $S_{850}^{\rm pred}/\rm mJy$ &  $S_{850}^{\rm obs}/\rm mJy$  \cr
\noalign{\phantom{A}}
2-449.0         &     23.39 &     23.68 &     23.92 &     25.72 &    0.32 &    2.85\rlap{${}^{*}$} &   \dots &    4.84 &    0.16 &    1.71 \cr
2-449.1         &     23.47 &     23.72 &     23.95 &     25.62 &    0.65 &    2.01\rlap{${}^{*}$} &    2.01\rlap{${}^{*}$} &    2.17 &    0.06 &    1.63 \cr
2-454.0         &     24.19 &     24.40 &     24.58 &     26.16 &    0.25 &    2.68 &    2.04 &    2.23 &    0.07 &    0.99 \cr
2-736.0         &     21.98 &     22.26 &     22.37 &     22.98 &    1.31 &    1.36\rlap{${}^{*}$} &   \dots &    3.50 &    0.09 &    1.94 \cr
2-736.11        &     22.94 &     23.11 &     23.16 &     23.71 &    1.07 &    1.62 &   \dots &    2.69 &    0.07 &    1.89 \cr
3-82.0          &     25.25 &     26.68 &     28.25 &     \dots &   \dots &    5.84 &    0.60 &    3.08 &    0.14 &   \min0.19 \cr
3-289.0         &     24.66 &     26.43 &     28.01 &     \dots &   \dots &    5.50 &    0.92 &    4.88 &    0.22 &    0.23 \cr
3-378.0         &     25.63 &     27.37 &     \dots &     \dots &   \dots &    5.62 &    1.00 &    2.06 &    0.09 &    0.04 \cr
3-571.0         &     24.79 &     25.92 &     27.26 &     \dots &   \dots &    5.11 &    0.72 &    3.91 &    0.17 &   \min0.09 \cr
3-839.0         &     24.66 &     26.82 &     28.92 &     \dots &   \dots &    5.39 &    2.08 &    4.75 &    0.21 &   \min0.32 \cr
3-951.0         &     25.58 &     28.06 &     \dots &     \dots &   \dots &    5.81 &   \dots &    2.26 &    0.11 &    0.85 \cr
3-965.1112      &     25.53 &     26.89 &     29.33 &     \dots &   \dots &    5.29 &   \dots &    2.07 &    0.09 &   \min0.97 \cr
4-439.1         &     24.96 &     26.03 &     \dots &     \dots &   \dots &    4.81 &    4.32 &    3.07 &    0.13 &    0.07 \cr
4-445.0         &     23.71 &     24.08 &     24.28 &     25.79 &    0.50 &    2.27\rlap{${}^{*}$} &    2.27\rlap{${}^{*}$} &    2.07 &    0.06 &   \min0.54 \cr
4-454.0         &     20.99 &     20.79 &     20.68 &     20.97 &    0.65 &    0.98 &   \dots &   10.00 &    0.26 &   \min2.37 \cr
4-555.1         &     23.17 &     23.41 &     24.10 &     \dots &    1.79 &    2.75 &   \dots &    5.21 &    0.17 &    1.28 \cr
4-555.12        &     23.94 &     24.21 &     24.98 &     \dots &    2.00 &    2.80\rlap{${}^{*}$} &    2.88 &    2.60 &    0.09 &    1.37 \cr
4-639.2         &     25.22 &     26.18 &     28.44 &     \dots &   \dots &    4.67 &    4.16 &    2.32 &    0.10 &    0.18 \cr
4-776.0         &     24.94 &     26.64 &     \dots &     \dots &   \dots &    5.53 &    0.96 &    3.80 &    0.17 &    0.53 \cr
4-878.11        &     23.22 &     23.53 &     23.93 &     25.48 &    1.10 &    2.33 &    0.00 &    3.24 &    0.10 &   \min0.44 \cr
}
}}

\endtable

\ssec{Comparison with NICMOS data}

To obtain accurate estimated redshifts at $z\gs 4$, 
infrared data are required. Part of the HDF (mainly chip WF4) has been imaged in
$J$$H$$K$ by Thompson et al. (1999). On the basis of the optical--infrared
colours, 8 candidates for galaxies at extreme redshifts were identified (Weymann et al. 2000),
including the $z=5.60$ object HDF4-473.0 (NICMOS source 184.0)
studied by Weymann et al. (1998).
The distribution of these objects is shown in Fig. 9, and their
parameters are listed in Table 2.
Although no significant sub-mm emission is associated with
HDF4-473.0, there is a striking association between the two
very close pairs of NICMOS objects and the two main 850-$\mum$ peaks
in the NICMOS field -- even extending to an agreement in position
angle between the Northernmost pair and the linear sub-mm source
mentioned earlier. Given the lack of
a plausible candidate identification for this object within the
optical HDF catalogue, the case for association with a $z\gs 5$ galaxy
seems rather compelling.

\japfig{9}{3}{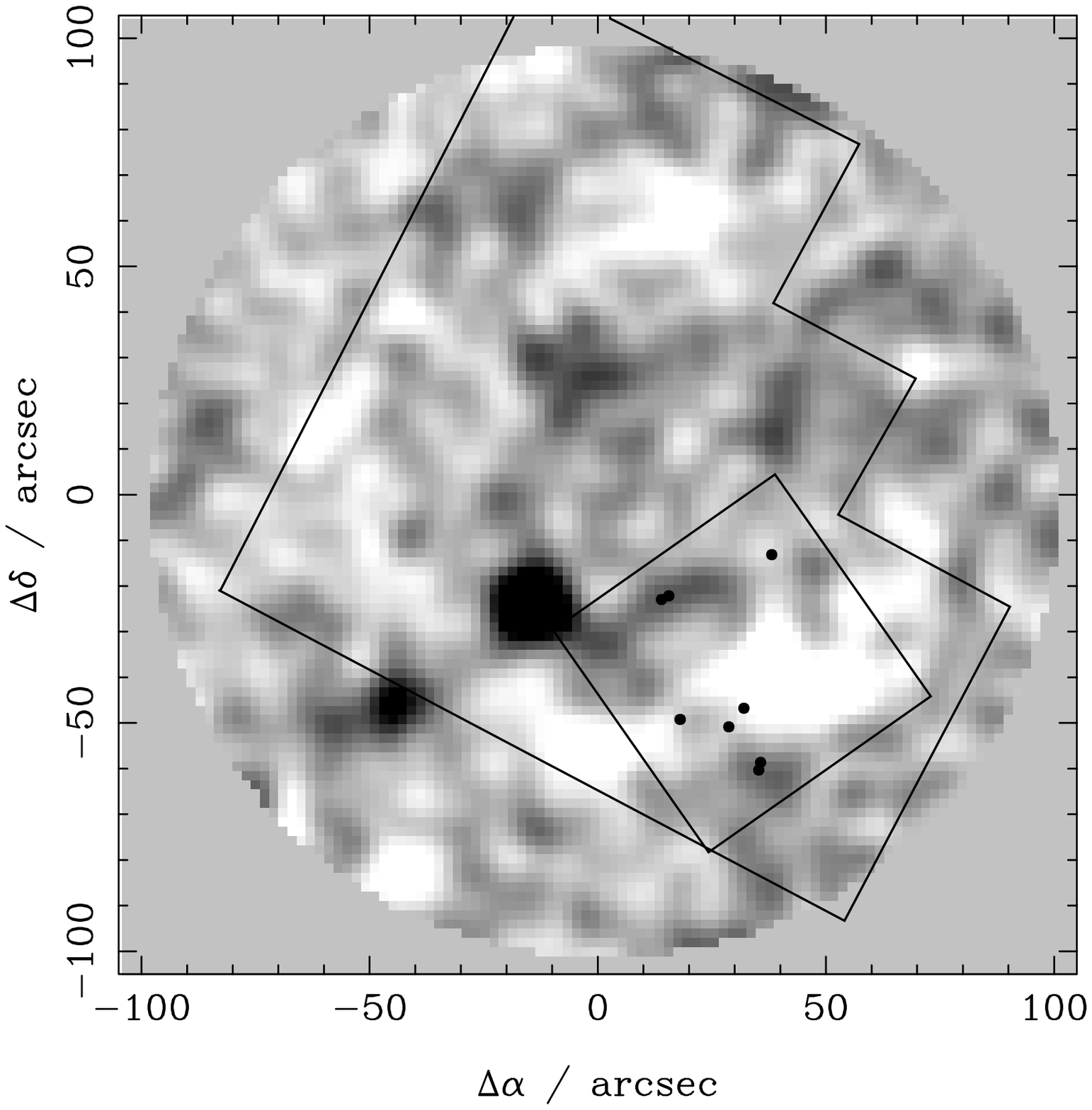}
{The NICMOS candidate ultra high-redshift galaxies
are shown as points. Interestingly, these appear to
coincide with weak SCUBA sources in the case of the
two close pairs.}

\begintable{2}
\caption{{\bf Table 2} The submillimetre properties of HDF/NICMOS
candidate high-redshift galaxies.}

\centerline{
\vbox{
\offinterlineskip
\tabskip 0.45cm
\halign{\phantom{\llap{$A^\beta_\beta$}}
\hfil# & \hfil#\hfil & \hfil#\hfil & \hfil#\hfil \cr
Name & $H_{160}$ & $J_{110}$ & $S_{850}^{\rm obs}/\rm mJy$  \cr
\noalign{\phantom{A}}
  96.0 &     28.81 &     28.51 &     \min0.15\cr
 118.0 &     27.86 &     28.65 &      1.88\cr
 123.0 &     28.08 &     \dots &      1.95\cr
 150.0 &     26.39 &     27.21 &     \min1.16\cr
 184.0 &     26.80 &     26.48 &     \min0.19\cr
 248.0 &     28.13 &     28.10 &      0.10\cr
 266.0 &     28.76 &     \dots &      1.27\cr
 678.0 &     27.93 &     27.60 &      1.35\cr
}
}}

\endtable

\sec{BACKGROUND CLUSTERING ANALYSIS}

\ssec{Angular power spectrum}

In order to improve on the somewhat anecdotal indications of
clustering in the 850-$\mum$ map, possibly associated with
high-redshift galaxies, it is necessary to quantify the degree
of anisotropy seen in the map. The most direct tool for this is
power-spectrum analysis, which allows a simple
decomposition of the various contributions to the map.
The map intensity, $I$, can be written as
$$
I = \left( S * W_1 + N \right) \times W_2,
$$
where $S$ is the true sky surface brightness; $W_1$ is the
`window function' corresponding to the instrumental
beam (including negative sidelobes from chopping \& nodding);
$N$ is the noise, which varies with radius;
$W_2$ is a `censor' function that sets the map to
zero beyond 100~arcsec radius, and is also chosen with
a radial variation to make the apparent noise independent
of radius.

The structure in the sky background emission can be described either
by its angular correlation function, $w(\theta)$, or by the
angular power spectrum, which is best written in dimensionless
form as $\Delta^2_\ell$, the fractional variance per logarithmic
increment of wavenumber:
$$
\eqalign{
&w(\theta) =\int_0^\infty \Delta^2_\ell\; J_0(\ell\theta)\; d\ell/\ell\cr
&\Delta^2_\ell =\ell^2\; \int_0^\infty w(\theta)\; J_0(\ell\theta)\;\theta\, d\theta.
}
$$
The angular wavenumber is written as multipole number, $\ell$, to emphasise
that the sky should really be expanded in spherical harmonics,
$Y_{\ell m}$. However, our survey subtends a small enough angle
that Fourier methods are an excellent approximation.
For power-law clustering, $w(\theta)=(\theta/\theta_0)^{-\epsilon}$,
this gives
$$
\Delta^2_\theta(\ell) = (\ell\theta_0)^\epsilon\; 2^{1-\epsilon}\;
{ \Gamma(1-\epsilon/2)\over\Gamma(\epsilon/2) },
$$
which is equal to $0.77(\ell\theta_0)^\epsilon$ for $\epsilon=0.8$.
This is the canonical slope for the small-separation
correlations of local galaxies, which seems to evolve remarkably
little with redshift (Giavalisco et al. 1998).

\ssec{Window functions}

An interesting aspect of the  850-$\mum$ HDF dataset is that it
contains a number of distinct scales, all of similar magnitude.
The small-scale instrumental noise has an effective FWHM of 8.5$''$
the telescope beam has a FWHM of 14.7$''$, and chop throws of
30$''$ and 45$''$; finally, the map radius is 100$''$, although
the noise doubles at radius 90$''$. Each of these numbers
controls the apparent clustering signal through an
appropriate window function.

The window for the beam is most easily evaluated as a two-step process.
Convolve the sky with a 14.7-arcsec FWHM Gaussian, then write the
signal as $S=S_3-(S_1+S_2+S_4+S_5)/4$, where (1,2,3,4,5) label
the positions in the chop/nod sequence. The variance for this
sum is easily written down in terms of the correlation function
of the smooth sky, which allows the window function to be identified:
$$
\eqalign{
|\tilde W_1|^2 &=
{5\over 4} + 
{1\over 4} J_0(\ell \theta) -
J_0(2\ell \theta) -
J_0(3\ell \theta) \cr
&+
{1\over 8} J_0(4\ell \theta) +
{1\over 4} J_0(5\ell \theta) +
{1\over 8} J_0(6\ell \theta),\cr
}
$$
where $\theta=15$~arcsec.

The noise in the published (slightly smoothed) map has an rms of 0.45~mJy in the
map centre, with a coherence length corresponding to filtering
with a Gaussian of FWHM 8.5~arcsec.
The spatial variation in the noise rms is divided out, so that
the censoring function is
$$
W_2=[1+(r/90\,{\rm arcsec})^2]^{-1}; \quad\quad r<R \equiv 100\,\rm arcsec.
$$
Without the radial variation, the $\ell$-space window would be just 
$\tilde W_2= \tilde W_3$, where
$$
\tilde W_3 = 2 J1(\ell R)/(\ell R);
$$
a good approximation to the effect of the
radial variation is to model $\tilde W_2$ by the
expression for $\tilde W_3$, but decreasing
the effective value of $R$ to 85~arcsec.

\ssec{Shot noise vs true clustering}

The power spectrum of the observed map is a sum of
three terms: instrumental noise, clustering of the
background galaxies, and shot noise arising because the
number of background galaxies is finite.

If all sources in the map had the same flux
density, the shot noise would be just
$$
\Delta^2_{\rm shot} = {\ell^2 \over 2\pi\, N},
$$
where $N$ is the surface density of sources. In the practical
case where the sources have different flux densities,
with number count $dN(S)$, the effective value of $N$ is
$$
N_{\rm eff} = { \left[ \int S\, dN(S) \right]^2 \over \int S^2\, dN(S) }.
$$
For Euclidean counts, $dN \propto S^{-5/2} dS$, and 
the shot noise diverges at high flux densities.
It therefore makes sense to work with the residual map where
the brighter point sources have been subtracted. This is
possible to a limit of 2~mJy in the 850-$\mum$ HDF map.

Blain et al. (1999) have shown that the number counts to
0.5~mJy are very close to Euclidean in form, with a normalization
of
$$
N(>1\,{\rm mJy}) = 7900 \pm 3000\; {\rm deg}^{-2}.
$$
If we assume that Euclidean counts with $dN=A S^{-5/2} dS$ hold 
between 2~mJy and some flux density
$S_0$, then
$$
N_{\rm eff} = {\sqrt{2} A\over S_0} (1-\sqrt{S_0/2}) 
$$
The background intensity is
$$
I = \int_{S_0}^\infty S\, dN(S) = 2A S_0^{-0.5}
$$
The observed background at 850~$\mum$ is
$$
\nu I=5\pm 2\times 10^{-10}\, {\rm W m^{-2}sr^{-1}} =
0.0033\,\rm mJy\, arcsec^{-2}
$$
(Fixsen et al. 1998), so the observed count
normalization ($A=11850$) requires $S_0=0.31$~mJy,
implying $N_{\rm eff} = 32800\; {\rm deg}^{-2}$.

It is unrealistic to assume abruptly truncated counts,
and indeed Blain et al. (1999) present evidence that the counts
have flattened by 0.5~mJy. Consider therefore the model
$$
{dN\over dS} = {A S^{-5/2} \over 1 + (S/S_0)^{-\gamma} },
$$
which is a more general variant of the form proposed by
Barger, Cowie \& Sanders (1999).
We have no direct constraints on $\gamma$, but comparison with the
faint cm-wavelength counts suggests $\gamma=1$, and we adopt this value.
The background is then $I=\pi A S_0^{-1/2}$, and
the surface density above 1~mJy is $2A S_0^{-3/2}[S_0^{1/2} - {\rm arctan}(S_0^{1/2})]$.
The observed counts and background then require $A=16500$,
$S_0=0.7$~mJy, and the effective shot-noise source density is
$N_{\rm eff} =92600\; {\rm deg}^{-2}$.
This is probably a more realistic estimate than the previous
figure; nevertheless, we have altered $N_{\rm eff}$ by a factor
of three through different models for the faint counts, so this number
has to be considered somewhat uncertain. Fortunately, it
can be estimated directly from the data, as shown below.

\japfig{10}{1}{2dpsa}{The two-dimensional power spectrum of the 850-$\mum$ HDF map.
Angular wavenumber is denoted by $\ell$, and $\Delta^2(\ell)$ is the
contribution to the fractional sky variance from unit range of $\ln\ell$.
The dotted line shows the expected noise spectrum; the
dot-dashed line is the shot-noise component for
an effective surface density of $N_{\rm eff}=92600\, \rm deg^{-2}$
(see text); the dashed line is the contribution from
clustering with $\theta_0=1$~arcsec; the solid line is the
total model power, which fits the data well.
}

\ssec{Results}

Fig. 10 shows the observed power spectrum estimated from the
850-$\mum$ map. The analysis requires a dimensionless
fluctuation field, so the map (after bright-source subtraction)
was divided by the expected background from sources at
$S<2$~mJy, which is 0.66 of the total, according to the above
count model, equivalent to a flux density of 0.54~mJy per
14.7-arcsec beam. The power spectrum was deduced
according to the methods described in Feldman, Kaiser \& Peacock (1994).
These authors showed that power estimates are correlated over
wavenumber separations of order the reciprocal of the survey
size, and this effect is noticeable for $\ell\ls 2\times 10^4$.

The data are compared with a model which is a sum of instrumental noise,
clustering, and shot noise.
These intrinsic effects are modified in two ways by instrumental
effects: the small-scale power is reduced by filtering with
the telescope beam (introducing a factor $|\tilde W_1|$), 
and the large-scale power is reduced
because the map is constrained to have zero mean flux
(Peacock \& Nicholson 1991); we cannot detect structures larger than the
map size, so the observed power must go to zero at
zero wavenumber (introducing a factor $[1-|\tilde W|^2]$).
The overall model is thus
$$
\eqalign{
\Delta^2_{\rm obs} &=
\left( \Delta^2_{\rm clus} + \Delta^2_{\rm shot} \right) \times 
|\tilde W_1|^2 \times (1-|\tilde W_2|^2) \cr
&+ \Delta^2_{\rm noise} \times (1-|\tilde W_3|^2) \cr
}
$$

Shot noise is distinguishable from true clustering because these effects
have a different dependence on scale. Our data cover only a restricted
range of scales, so in practice we shall assume that the true
angular clustering has a correlation structure of the same shape
as is observed for most faint galaxy samples, but with an
unknown amplitude: $w(\theta)=(\theta/\theta_0)^{-0.8}$.

\japfig{11}{1}{clus_poiss}
{Contours of likelihood relative to maximum for the fit of
the 850-$\mum$ power spectrum data by a combination of
shot noise (effective surface density for sources
with flux density $<2$~mJy $N_{\rm eff}$) and
galaxy clustering (where $\theta_0$ is the scale-length
in the angular correlation function).
The contour interval is 0.5 in $\ln{\cal L}$,
corresponding to confidence contours of approximately
39\%, 63\%, 78\%, 87\%, 92\% etc.}

Fig. 10 shows an illustrative fit of shot noise with 
$N_{\rm eff} =92600\; {\rm deg}^{-2}$ as estimated above,
and clustering with $\theta_0=1$~arcsec. However,
both these parameters can be fitted to the data in the
usual maximum likelihood manner. The result of this
exercise is shown in Fig. 11. As would be expected,
there is a degree of degeneracy in the free parameters;
the data indicate an excess of power in the residual
map above pure noise for $10^4 \ls \ell \ls 5\times 10^4$,
and this constrains the sum of clustering and shot power.
The different scale dependencies of the two signals
mean that the degeneracy is broken, and the
preferred model has significant clustering
($\theta_0=3$~arcsec) and low shot noise
($N_{\rm eff} = 5\times 10^5\, {\rm deg}^{-2}$).
However, this is an unrealistically high
surface density; we should really fix $N_{\rm eff}$ at its a priori estimated
value, in which case there is no significant detection of
clustering, and $\theta_0=3$~arcsec can be ruled out at
approximately 95\% confidence.

\ssec{Implications for origin of the background}

What is the interpretation of the relatively weak angular clustering?
The angular power is a projection of the true 3D spatial
power spectrum, given by the k-space version of 
Limber's equation:
$$
\Delta^2_\theta={\pi\over \ell}\int\Delta^2(\ell/y)\; C(y) y^5\phi^2(y)\; dy
$$
(Kaiser 1992; see section 16.6 of Peacock 1999). 
In this expression, $y$ is comoving angular-diameter distance, and the
`selection function' $\phi(y)$ is normalized so that
$$
\int_0^\infty y^2\phi(y) C(y) dy=1.
$$
The curvature factor 
$$
C(y)=\left[ 1 + {(1-\Omega) y \over c/H_0} \right]^{-1/2},
$$
so it is unity for spatially flat models.

If $\Delta^2(k) \propto k^{1.8}$ (so that $\Delta^2_\theta(\ell) \propto \ell^{0.8}$),
the angular power scales as
$$
\Delta^2_\theta \propto  {
\int y^{3.2} \phi^2(y) C(y)\; dy
\over
\left[ \int  y^2\phi(y)C(y)\; dy \right]^2
}
$$
The redshift distribution relates to the selection function via
$n(z) dz \propto y^2\phi(y) C(y) dy$. What this says is that the
amplitude of angular clustering depends on the range of redshifts
that contribute to the sample under study, and a low
value of $\theta_0$ requires a large range of
redshifts to be contributing. 

To be quantitative, consider the results of Giavalisco et al. (1998)
for the angular clustering of Lyman-break galaxies.
They deduce $\theta_0=2$~arcsec over a redshift range
$2.5 \ls z \ls 3.5$.
The relation of comoving distance to observables is
$$
\eqalign{
&C(y)\, dy = {c\over H_0} \, dz \;\;\times \cr
&\;\;\;\; \left[ (1-\Omega_m-\Omega_v)(1+z)^2
+\Omega_v + \Omega_m(1+z)^3\right]^{-1/2},\cr
}
$$
and this allows us to see what would happen to the angular
clustering signal if the redshift range is extended. Assuming
a uniform contribution to the sub-mm surface brightness
from $2 < z < 6$, as indicated by the results in earlier
sections, the angular clustering signal would be expected to
drop by a factor of 0.35 ($\Omega_m=1$, $\Omega_v=0$)
or 0.34 ($\Omega_m=0.3$, $\Omega_v=0.7$) compared to that
measured by Giavalisco et al., for the same degree
of spatial clustering. 

In making this comparison, it is important to be clear that
we are including the angular clustering of the ULIRG
population, rather than just that of Lyman-break galaxies,
since the former give a significant contribution to the background.
Nevertheless, the Lyman-break galaxies serve as a useful
reference. They are a highly biased population, as indeed
would be expected for all reasonably massive galaxies at
high redshift (Steidel et al. 1998; Peacock et al. 1998).
It would therefore not be surprising if the ULIRG population
displayed a similar level of spatial clustering.
However, even if this is so, the
sub-mm background arises over a broad range of
redshifts, and the sky at 850-$\mum$ is therefore 
expected to be more uniform than the Lyman-break sky.
Strong clustering of the 850-$\mum$ background would
only arise if the ULIRGs were much more strongly clustered
that Lyman-break galaxies; our results indicate that this cannot
be the case.

As a final point, we can now address the question of whether
clustering in the background could bias the statistical detection of 
flux from the UV starburst galaxies. Any given galaxy will
statistically  be surrounded by a `halo' of emission contributed
by correlated galaxies, whose surface brightness at
angular separation $\theta$ will be $w(\theta) I$, where $I$
is the mean background. Since we are dealing with two
populations, UV starbursts and ULIRGs, in principle we need both
the UV--ULIRG cross-correlation function as well as the UV autocorrelation
function. However, since a correlation coefficient must be
less than unity, the cross-correlation cannot exceed the larger
of the two autocorrelations. We can therefore get a limit
to the effect by assuming $\theta_0=2$~arcsec. The effective
point-source flux is obtained by integrating $w(\theta) I_\nu$ over
the beam. The result of this exercise is an effective point-source
flux density of 0.19~mJy. This is large enough to be a dominant
bias for the starbursts with SFR around $1\msunyr$, but 
cannot erase the signal from the more luminous objects.
In order to obtain an upper limit to the overall bias that could
be introduced by this effect, the analysis of section 3.3 for the
mean level of 850-$\mum$ emission was repeated, subtracting 0.19~mJy
from all the flux densities. The result is to reduce the mean
sub-mm output for a given SFR by a factor of approximately 2,
although we have argued that
this is very much an upper limit to the correction.
The effect of clustering can therefore in principle have a significant
quantitative impact on issues such as the total sub-mm emission from
UV starbursts (section 3.4). However, this does not alter the
main point, which is that UV starbursts are one of the largest
contributors to the background radiation at these frequencies.

\sec{CONCLUSIONS}

In this paper, we have investigated whether the SCUBA 850-$\mum$ map of
the HDF can set useful constraints on the general 
population of optically-selected starburst
galaxies, or whether we are confined to investigating only totally
obscured ultraluminous infrared galaxies. 
Although data of improved resolution and sensitivity will be
required in order to study the sub-mm properties of HDF galaxies
on an individual basis, we believe that there is good statistical
evidence that the galaxies detected in the optical HDF catalogues
are emitting in the sub-mm band, 
with a flux density of about $S_{850}=0.2$~mJy for 
an apparent UV star-formation rate of $1\msunyr$.
This level of emission
is consistent with the idea that the UV emission from these 
galaxies underestimates the total star-formation rate in most cases,
by a mean factor of approximately 6. This means that the
Lyman-break population must contribute at least 25\% of
the background radiation at 850~$\mum$.

We have used a variety of tests, including cross-correlation
with UV data and power-spectrum analysis of the 850-$\mum$ data,
to demonstrate that the sub-mm background receives approximately 
equal contributions from a very wide range of redshifts,
$1\ls z \ls 6$. Indeed, some of the strongest individual
examples of evidence for an association between high-z
galaxies and features in the sub-mm map involve galaxies towards
the higher end of this range. 
Together with the lack of
a clear optical counterpart for the brightest HDF
source (HDF 850.1; Downes et al. 1999), this suggests
that the upper bound of the redshift distribution for
850-$\mum$ sources may not yet have been reached.

\section*{Acknowledgements}

We thank Len Cowie and especially Steve Eales for useful
comments on this work.
The James Clerk Maxwell Telescope is operated by the Joint Astronomy Centre
on behalf of the Particle Physics \& Astronomy Research Council of the
United Kingdom, The Netherlands Organisation for Scientific Research, and
the National Research Council of Canada.

\section*{References}

{

\pretolerance 10000

\ref Almaini O., Lawrence A., Boyle B.J., 1999, \mn, 305, L59
\ref Barger A.J., Cowie L.L., Sanders D.B., 1999, \apj, 518, L5
\ref Barger A.J., Cowie L.L., Smail I., Ivison R.J., Blain A.W., Kneib J.-P., 1999, \aj, 117, 2656
\ref Blain A.W., Kneib J.-P., Ivison R.J., Smail I., 1999, \apj, 512, L87
\ref Busswell G.S., Shanks T., 2000, astro-ph/0002081 
\ref Chapman S.C. et al., 1999, astro-ph/9909092
\ref Downes D., et al., 1999, A\&A,  347, 809
\ref Eales S., Lilly S., Gear W., Dunne L., Bond J.R., Hammer F., Le F\'evre O., Crampton D., 1999, \apj, 515, 518
\ref Feldman H.A., Kaiser N., Peacock J.A., 1994, \apj, 426, 23
\ref Fernandez-Soto A., Lanzetta K.M., Yahil A., 1999, \apj, 513, 34
\ref Fixsen D.J., Dwek E., Mather J.C., Bennett C.L., Shafer R. A., 1998, \apj, 508, 123
\ref Giavalisco M., Steidel C.C., Adelberger K.L., Dickinson M.E., Pettini M., Kellogg M., 1998, \apj, 503, 543
\ref Gunn K.F., Shanks T., 1999, astro-ph/9909089
\ref Hauser M.G. et al., 1998, \apj, 508, 25
\ref Holland W.S., et al., 1999, MNRAS, 303, 659
\ref Hogg D.W. et al., 1998, AJ, 115, 1418
\ref Hughes D., Dunlop J.S., 1999, in {\it Highly Redshifted Radio Lines}, eds  C.L. Carilli, S.J.E. Radford, K.M. Menten, G.I. Langston, ASP Conf. Ser. {\bf 156}, p99 (astro-ph/9802260)
\ref Hughes D. et al., 1998, Nature, 394, 241
\ref Ivison R.J., Smail I., Barger A.J., Kneib J.-P., Blain A.W., Owen F.N., Kerr T., Cowie L.L., 1999, astro-ph/9911069
\ref Jimenez R.,  Padoan P., Dunlop J., Bowen D., Juvela M., Matteucci F., 2000, \apj, 532, 152
\ref Kaiser N., 1992,  \apj, 388, 272
\ref Lilly S.J., Le F\`evre O., Hammer F., Crampton D., 1996, \apj, 460, L1
\ref Lilly S.J., Eales S.A., Gear W.K.P., Hammer F., Le F\`evre O., Crampton D., Bond J.R., Dunne L., 1999, \apj, 518, 641
\ref Madau P. et al., 1996, MNRAS, 283, 1388
\ref Meurer G.R., Heckman T.M., Lehnert M., Leitherer C., Lowenthal J., 1997, \aj, 114, 54
\ref Meurer G.R., Heckman T.M., Calzetti D., 1999, \apj, 521, 64
\ref Peacock J.A., Nicholson D., 1991, MNRAS, 253, 307
\ref Peacock J.A., Jimenez R., Dunlop J.S., Waddington I., Spinrad H., Stern D., Dey A., Windhorst R.A., 1998, \mn, 296, 1089
\ref Peacock J.A., 1999, {\it Cosmological Physics}, Cambridge University Press
\ref Pettini M., Kellogg M. Steidel C.C., Dickinson M., Adelberger K.L., Giavalisco M., 1998, \apj, 508, 539
\ref Puget J.-L., Abergel A., Bernard J.-P., Boulanger F., Burton W.B., Desert F.-X., Hartmann D., 1996, A\&A, 308, L5
\ref Rowan-Robinson M., 2000, in preparation.
\ref Schlegel D.J., Finkbeiner D.P., Davis M., 1998, \apj, 500, 525
\ref Smail I., Ivison R.J., Blain A.W., 1997, \apj, 490, L5
\ref Smail I., Ivison R.J., Kneib J.-P., Cowie L.L., Blain A.W., Barger A.J., Owen F.N., Morrison G.E., 1999, \mn, 308, 1061
\ref Smail I., Ivison R.J., Owen F.N., Blain A.W., Kneib J.-P., 2000, \apj, 528, 612
\ref Soifer B.T., Sanders D.B., Madore B.F., Neugebauer G., Danielson G.E., Elias J.H., Lonsdale C.J., Rice W.L., 1987, \apj, 320, 238
\ref Steidel C.C., Giavalisco M., Pettini M., Dickinson M., Adelberger K.L., 1996, \apj, 462, L17
\ref Steidel C.C., Adelberger K.L., Dickinson M., Giavalisco M., Pettini M., Kellogg M., 1998, ApJ, 492, 428
\ref Steidel C.C., Adelberger K.L., Giavalisco M., Dickinson M., Pettini M., 1999, \apj, 519, 1
\ref Thompson R.I., Storrie-Lombardi L.J., Weymann R.J., Rieke M.J., Schneider G., Stobie E., Lytle D., 1999, AJ, 117, 17
\ref Weymann R.J., Stern D., Bunker A., Spinrad H., Chaffee F.H., Thompson R.I., Storrie-Lombardi L.J., 1998, \apj, 505, L95
\ref Weymann R.J. et al., 2000, in preparation.
\ref Williams R.E. et al., 1996, AJ, 112, 1335

}

\bye